\documentclass{PoS}
\title{The density of states from first principles}

\ShortTitle{The density of states from first principles}

\author{\speaker{Roberto Pellegrini}\\
        Department of Physics, Swansea University, Singleton Park, Swansea, SA2 8PP UK\\
        E-mail: \email{pyrp@swansea.ac.uk}}

\author{K. Langfeld\\
        School of Computing and Mathematics, Plymouth, PL4 8AA, UK\\
        E-mail: \email{kurt.langfeld@plymouth.ac.uk}}

\author{B. Lucini\\
        Department of Physics, Swansea University, Singleton Park, Swansea, SA2 8PP UK\\
        E-mail: \email{b.lucini@swansea.ac.uk}}

\author{A. Rago\\
        School of Computing and Mathematics, Plymouth, PL4 8AA, UK\\
        E-mail: \email{antonio.rago@plymouth.ac.uk}}

\abstract{We present a novel algorithm to compute the density of states, which is proven to converge to the correct result. The algorithm is very general and can be applied to a wide range of models, in the frameworks of Statistical Mechanics and Lattice Gauge Theory. All the thermal or quantum expectation values can then be obtained by a simple integration of the density of states. As an application, a numerical study of $4d$ U(1) compact lattice gauge theory is presented.}

\FullConference{The 32nd International Symposium on Lattice Field Theory,\\
		23-28 June, 2014\\
		Columbia University New York, NY}

\begin{document}

\section{The density of states}
Monte-Carlo methods are one of the most developed tools to study non-perturbatively quantum field theories from first principles. 
This approach is very powerful and delivers high precision results. However its applicability is limited to the case of
observables that can be expressed as vacuum expectation values (vev) over a probability measure. Partition functions are an example
of observables that are not a vev, while at finite density the path integral measure of QCD cannot be interpreted as a
probability measure. In these cases Monte-Carlo methods are either unsuitable or very inefficient.
We propose an alternative method based on the density of states that is free from the described limitations.
Let us consider an euclidean quantum field theory and its partition function
\begin{equation}
  Z(\beta)=\int [ D \phi ] e^{-\beta S [ \phi ]},
\end{equation}
where $ \phi $ denotes the fields and $\int [ D \phi ]$ is the path integral over the field configurations.  
We define the density of states with the following functional integral
\begin{equation}
  \rho(E)=\int [ D \phi ] \delta(S[\phi]-E).
\end{equation}
From the density of state we can recover the partition function via a simple one dimensional
integral
\begin{equation}
 Z(\beta)=\int dE \rho(E) e^{-\beta E}
\end{equation}
and expectation values of observables $O(E)$ from
\begin{equation}
 \langle O(E) \rangle=\frac{1}{Z}\int dE O(E) \rho(E) e^{-\beta E}. 
\end{equation}
\section{A novel algorithm for the density of states}
Wang and Landau proposed a numerical algorithm to compute the density of states
in systems with discrete energy levels \cite{WL}. The algorithm is very well known in the Statistical Mechanics community
and has been an important pillar for obtaining remarkable results that would not have been possible to achieve using standard Monte-Carlo. 
Despite its success in the Statistical Mechanics framework it has found very limited applications
in Lattice Gauge Theory, the main reason being that a direct generalization to continuous systems seems not to be very efficient
\cite{ContWangLand1,ContWangLand2}.
We propose a different algorithm based on \cite{LLR} suitable for continuous systems.
The underlying assumption is that if we consider a small energy interval $[E_0-\Delta,E_0+\Delta]$ the logarithm of the density of states
can be well approximated by a piecewise linear function
\begin{equation}
 \rho(E)=e^{a(E_0)(E-E_0)}+ \mathcal{O}(\Delta) \label{linear}.
\end{equation}
The purpose of the algorithm is to calculate the coefficients $ a(E) $, which are related to the density of states via
\begin{equation}
 a(E_0)=\left. \frac{d \log(\rho(E))}{dE} \right|_{E=E_0}.
\end{equation}
Let us consider the expectation value of a function $f(E)$ restricted to the energy interval $[E_0-\Delta,E_0+\Delta]$ which we denote
with double angular brackets
\begin{equation}
 \langle \langle f(E) \rangle \rangle_a=\frac{1}{Z}\int_{E_0-\Delta}^{E_0+\Delta} f(E) \rho(E) e^{-a E} dE.
\end{equation}
We can choose $f(E)=E-E_0$, which gives
\begin{equation}
 \Delta E(a)=\langle \langle E-E_0 \rangle \rangle_a=\frac{1}{Z}\int_{E_0-\Delta}^{E_0+\Delta} (E-E_0) \rho(E) e^{-a E} dE. \label{rev}
\end{equation}
If eq.~(\ref{linear}) is a good approximation we should be able to choose $a^{*}$ such that
\begin{equation}
 \rho(E)exp(-a^{*} E)=constant + \mathcal{O}(\Delta)
\end{equation}
which inserted in eq.~(\ref{rev}) yields
\begin{equation}
 \Delta E(a^*)=0. \label{rfp}
\end{equation}
The truncated expectation value $\Delta E(a) $ can be computed by means of a standard Monte-Carlo
procedure that samples the configurations with weight
\begin{equation}
W(E) \propto e^{-a E} (\theta(E-E_{0}+\Delta)-\theta(E-E_{0}-\Delta)). \label{weight}
\end{equation}
Our goal is to find $ a^{*} $ such that this Monte-Carlo procedure gives eq.~(\ref{rfp}), which is a root finding
problem. 
There are many numerical root finding algorithms that can be found in textbooks. One of the most used 
is the iterative Newton-Raphson method. This method starts with a guessed value $a_0$ which is then updated according to the
iteration
\begin{equation}
 a_{n+1}=a_{n}-\frac{\Delta E(a_{n})}{\Delta E(a_{n})^{'}} \sim a_{n}-\frac{3 \Delta E (a_{n})}{\Delta^{2}}. \label{NR}
\end{equation}
Were $\Delta E(a)$ a deterministic and sufficiently well-behaved function then $ a_n $ would converge
to the true root $a^{*}$
\begin{equation}
 \lim_{n \rightarrow \infty} a_n=a^{*}.
\end{equation}
$ \Delta E(a) $ in our case is obtained by a Monte-Carlo procedure and thus it is not a deterministic
function. While the method seems to give the correct answer \cite{LLR,Kurt1,Biagio1}, the deterministic nature of the function $\Delta E(a)$ is a necessary
hypothesis for the mathematical proof of convergence.

In 1951 Robbins and Monro presented a different algorithm for solving root finding problems \cite{RM} where the
function can be expressed as an expectation value.
In other words we can drop the deterministic hypothesis and in place of the latter we ask
that we can obtain measurements of the function $ m(a)$ such that
\begin{equation}
 \Delta E(a)=\langle m(a) \rangle.
\end{equation}
The Robbins Monro iteration is given by
\begin{equation}
 a_{n+1}=a_{n}- c_{n} \Delta E(a_n),
\end{equation}
where
\begin{equation}
\sum_{n=0}^{\infty} c_{n}= \infty; \ \ \sum_{n=0}^{\infty} c_{n}^{2} < \infty.
\end{equation}
Under some mild assumptions on the function Robbins and Monro proved that $\lim_{n \rightarrow \infty}a_{n} $
converges to the true value $ a^{*} $ almost surely and thus also in probability.

A major advance in understanding the asymptotic properties of this algorithm was the main result of \cite{Sacks}.
Following the author we restrict ourselves to the case
\begin{equation}
 c_n=\frac{c}{n};
\end{equation}
it is then possible to prove that $ \sqrt{n}(a_{n}-a_{*}) $ is asymptotically normal
with variance
\begin{equation}
\sigma_{a}^{2}=\frac{c^2 \sigma_{\xi}^{2}}{2c f^{'}(a_{*})-1}, \label{var}
\end{equation}
where $\sigma_{\xi}^2$ is the variance of the noise.

In this sense the optimal value of the constant $c$ is
\begin{equation}
c=\frac{1}{f^{'}(a_{*})}. 
\end{equation}
In our particular case we have
\begin{equation}
 \Delta E(a^{*})^{'}=\frac{\Delta^2}{3};
\end{equation}
consequently we adopted the following iteration
\begin{equation}
 a_{n+1}=a_{n}- \frac{3 \Delta E(a_n)}{(n+1) \Delta^{2}},
\end{equation}
which is proved to converge to the correct value $a^{*}$ and optimal in the sense
that it minimizes the variance in eq.~(\ref{var}).
\section{Ergodicity}
\begin{figure}
\begin{center}
 \includegraphics[scale=0.6]{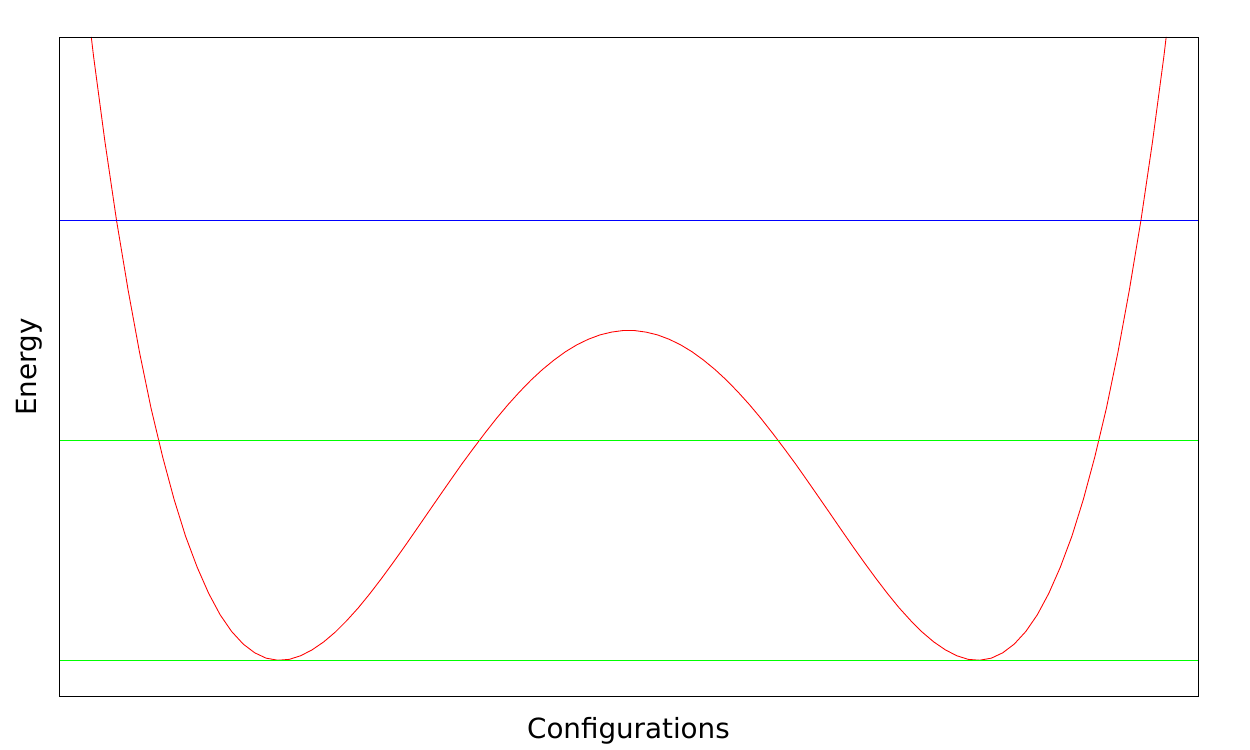}
\caption{\label{rugged} In red a hypothetical energy landscape, in this case the algorithm is trapped in one of the two minima.}
\end{center}
\end{figure}

\begin{figure}
\begin{center}
 \includegraphics[scale=0.6]{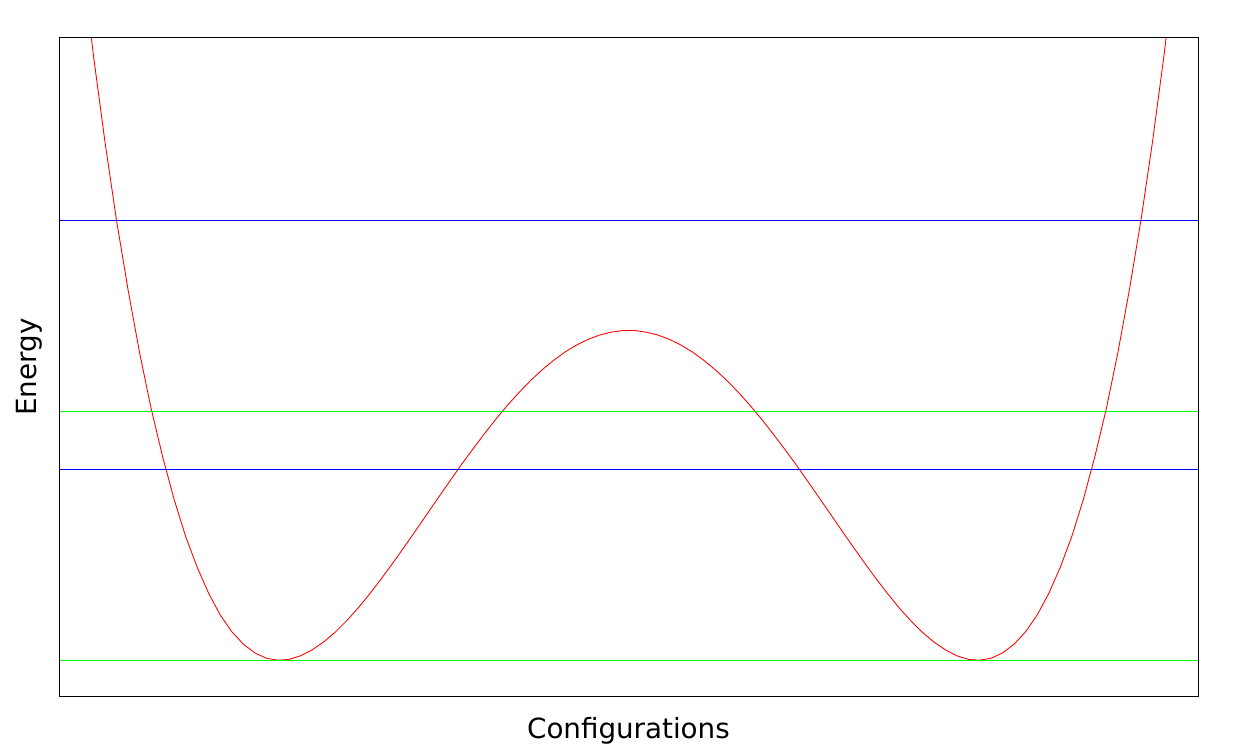}
\caption{\label{parallel} Using overlapping energy intervals the algorithm is allowed to sample
the whole phase space.}
\end{center}
\end{figure}

Using the weight in eq.~(\ref{weight}) we are forcing the system in an energy interval and
if the energy landscape is rugged we are not exploring all the configurations in an energy interval see fig.~(\ref{rugged}).
A possible solution, in analogy with parallel tempering, has been proposed recently
for the Wang-Landau algorithm \cite{ParallelWang}. We can consider overlapping energy intervals (see fig.~(\ref{parallel})) and simulate these
intervals in parallel. When two simultaneous simulations are in the overlapping region of two different
energy intervals we swap the systems with probability
\begin{equation}
P_{sw} = min(1, exp(\Delta a \Delta E )),
\end{equation}
where $\Delta E $ and $\Delta a$ are the differences of energies and of the coefficients $a$ of the two runs at a set number of iterations.  
This version of the algorithm satisfies the detailed balance condition and samples the whole phase space
of the theory.
\section{Test in 4d $U(1)$ gauge theory}
\begin{figure}
\begin{center}
 \includegraphics[scale=0.6]{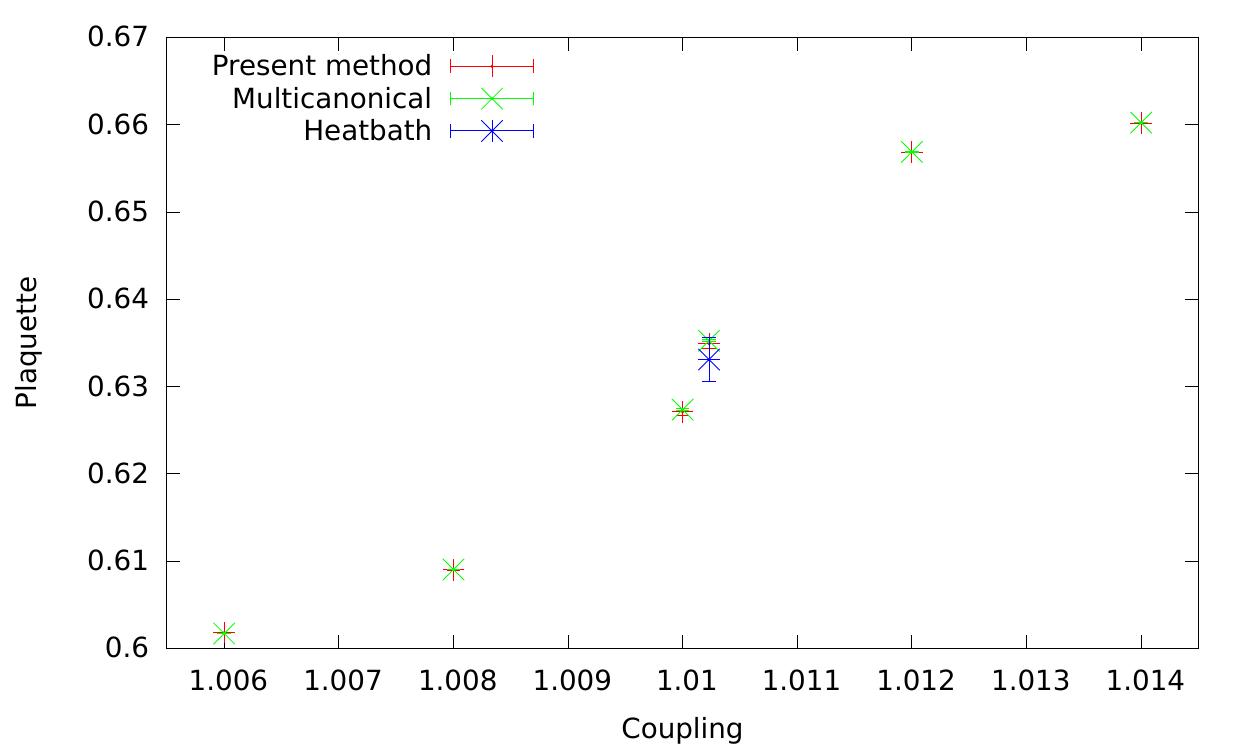}
\caption{\label{plaq} Plaquette values using different methods for $12^4$ lattice. }
\end{center}
\end{figure}

We tested our approach in the $4d$ compact $U(1)$ gauge theory, with
Wilson action. Here the link variables are points of the complex unit circle, i.e.,
$U_{\mu}(x)=exp(i \theta_{\mu}(x))$ with $\theta_{\mu}(x) \in [-\pi,\pi]$.
As a simple check we computed the plaquette in a $12^4$ lattice using our method, a multi-canonical simulation \cite{Berg}
and a standard heat-bath algorithm; the results are reported in fig.(\ref{plaq}) and they
are in good agreement with each other.
For this run we computed the density of states $\rho(E)$ for values of $E$ in $ [0.59,0.69] $ and we divided this
interval in 512 subintervals. In each interval, the errors are determined by a boostrap analysis of 20 independent simulations.
The whole computation required around $ 512*18 $ hours of CPU time. 

\begin{figure}
\begin{center}
 \includegraphics[scale=0.6]{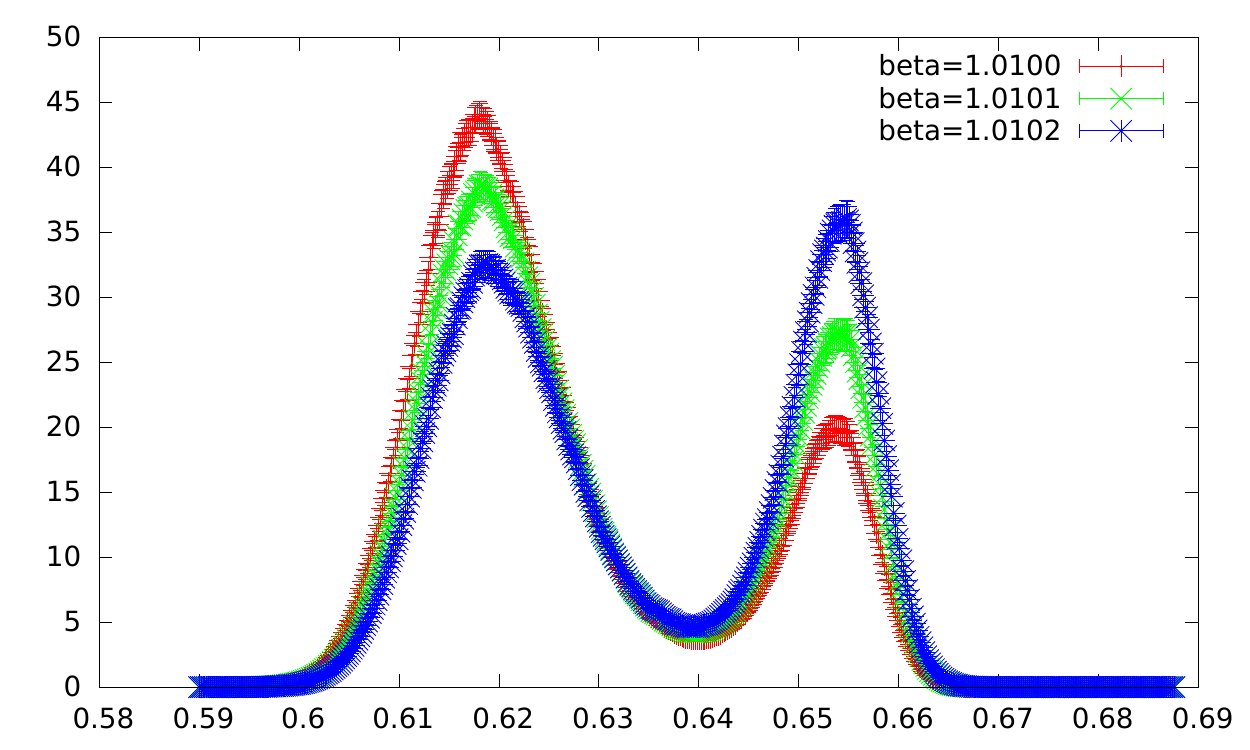}
\caption{\label{prob} $P(E)_{\beta}$ for different values of $\beta$.}
\end{center}
\end{figure}

By means of a large scale investigation on the basis of the Borgs-Kotecky finite size scaling analysis,
it has been established that the system undergoes a very weak first order phase transition at $\beta_c=10111331(21)$ \cite{Arnold1}.
A sign of this transition is the double peaks structure in the action probability density
\begin{equation}
 P(E)_{\beta}=\frac{1}{Z} \rho(E)exp(\beta E),
\end{equation}
which is immediately visible in fig.(\ref{prob}).
We also located the pseudo-critical coupling using the peak of the specific heat for volumes from $8^4 $ to $ 20^4 $
and we found a good agreement with \cite{Arnold2}, see tab.(\ref{peak}).
\begin{table}
\begin{center}
\begin{tabular}{c  c  c }
  L & Cv peak present method & Cv peak from \cite{Arnold2}  \\
  \hline
  8  & 1.00744(2) & 1.00741(1)  \\
  10 & 1.00939(2) & 1.00938(2)\\
  12 & 1.01025(1) & 1.01023(1)\\
  14 & 1.010624(5)& 1.01063(1)\\
  16 & 1.010833(4)& 1.01084(1)\\
  18 & 1.010948(2)& 1.010943(8)\\
  20 & 1.011006(2)&  \\
  \end{tabular} 
  \caption{Comparison between the location of the specific heat computed with the proposed method
  and the values in \cite{Arnold2}. The longest run ($20^4$ lattice) required $512*144$ hours of CPU time.}\label{peak}
\end{center}
  \end{table}

\section{Discussion and conclusion}
In this contribution we presented a novel algorithm to compute the density of states from first principles in continuous system.
The algorithm is mathematically solid and gives correct results in the limit of very small energy intervals.
There are many potential applications for which this algorithm is particularly suitable such as first principles studies of monopoles, vortices and interfaces.
Using the density of states it is also possible to transform a high dimensional oscillatory problem (finite density theories) to a single dimensional oscillatory integral
(see \cite{Biagio2}). Future directions include feasibility studies in theories with fermions and a numerical analysis of the effects due to the finite width of the energy intervals.
\section*{Acknowledgement}
We thank L. Bongiovanni for useful discussion. We are grateful for the support from the High Performance Computing Wales, where the numerical computations have
been carried out. KL and AR are supported by Leverhulme Trust (grant RPG-2014-118) and STFC (grant ST/L000350/1). BL and RP are supported by STFC (grant ST/G000506/1).

%\fi
\end{document}